\documentclass[12pt]{iopart}
\newcommand{\be}{\begin{equation}}
\newcommand{\ee}{\end{equation}}
\newcommand{\bea}{\begin{eqnarray}}
\newcommand{\eea}{\end{eqnarray}}
\usepackage{graphicx,psfig,iopams}  
\begin{document}

\title[Effects of network topology on wealth distributions]{Effects of network topology on wealth distributions}

\author{Diego Garlaschelli$^1$ and Maria I Loffredo$^2$}
\address{$^1$ Dipartimento di Fisica, Universit\`a di Siena, Via Roma 56, 53100 Siena ITALY.}
\ead{garlaschelli@unisi.it}
\address{$^2$ Dipartimento di Scienze Matematiche ed Informatiche, Universit\`a di Siena, Pian dei Mantellini 44, 53100 Siena ITALY.}
\ead{loffredo@unisi.it}

\begin{abstract}
We focus on the problem of how wealth is distributed among the units of a networked economic system. We first review the empirical results documenting that in many economies the wealth distribution is described by a combination of log--normal and power--law behaviours. We then focus on the Bouchaud--M\'ezard model of wealth exchange, describing an economy of interacting agents connected through an exchange network. We report analytical and numerical results showing that the system self--organises towards a stationary state whose associated wealth distribution depends crucially on the underlying interaction network. In particular we show that if the network displays a homogeneous density of links, the wealth distribution displays either the log--normal or the power--law form. This means that the first--order topological properties alone (such as the scale--free property) are not enough to explain the emergence of the empirically observed \emph{mixed} form of the wealth distribution. In order to reproduce this nontrivial pattern, the network has to be heterogeneously divided into regions with variable density of links. We show new results detailing how this effect is related to the higher--order correlation properties of the underlying network. In particular, we analyse assortativity by degree and the pairwise wealth correlations, and discuss the effects that these properties have on each other.

\end{abstract}


\section{Introduction}
Real economies are an example of human complexity at the largest scale, where the local interactions of individuals result in globally organised properties of the socioeconomic system as a whole. 
One of the most important collective properties of an economy is the distribution of the wealth owned by its fundamental units  \cite{sizes}. 
While the understanding of this process has traditionally invoked dynamical mechanisms treating each economic agent as a separate entity, more recently the role of the interactions has been considered as a key element in the self--organisation of the system  \cite{econowealth,yako}. Making use of the results on complex networks \cite{caldarelli,cosinbook}, it is now possible to consider interaction networks explicitly and understand the effects that their topological properties have on wealth dynamics.\\

In this work, we first review some stylized facts characterizing empirical wealth distributions  \cite{sizes,econowealth,yako,pareto,mandelbrot,atkinson,souma0,souma1,fabietto,tiziana}. Then we focus on the role of topology in determining how wealth redistributes among the units of a networked economic system. By means of a stochastic multi--agent model  \cite{bouchaud}, we confirm previous results  \cite{souma2,souma3,mypavia} showing that either log--normal or power--law distributions arise when the network is homogeneous (with uniform density of connections). 
As a consequence, we find that in order to obtain the most general observed pattern characterized by a combination of both distributions, different regions of the network have to be heterogeneously connected. We finally relate these results to higher--order correlation properties such as assortative mixing by wealth and degree.

\section{Economic distributions: empirical results}
Many empirical studies  \cite{sizes,econowealth,yako,pareto,mandelbrot,atkinson,souma0,souma1,fabietto,tiziana} 
report universal properties of wealth  and income distributions, even if quantitative differences exist across various economies. 
One common feature, first pointed out by the early work of the Italian economist and social scientist Vilfredo Pareto  \cite{pareto}, is that the large wealth range is empirically found to be power--law distributed, i.e. the statistical distribution of the wealth $w$ has the form $p(w)\sim w^{-\mu}$ for $w\to\infty$, $\mu>0$. The exponent of the power--law is usually written in the form $\mu=1+\alpha$, where $\alpha$ has to be positive for $p(w)$ to be integrable in the large wealth limit. In real economies, $\alpha$ (also called the \emph{Pareto index}) is found to be time--dependent within the range $1\le\alpha(t)\le 2.5$  \cite{pareto,mandelbrot,souma1}. To avoid divergence at zero wealth, the distribution must hold only above a threshold value $w_*>0$. The correctly normalized \textbf{Pareto's law} is then
\be
\label{pareto}
p(w)=\frac{\alpha w_*^\alpha}{w^{1+\alpha}}\qquad w\ge w_*
\ee
and $p(w)=0$ for $w<w_*$. Note that the cumulative distribution $P_>(w)\equiv\int_w^\infty p(w')dw'$ behaves as a power law with exponent $-\alpha$. 

Power laws lack a characteristic scale, and are related to fractal structures with self--similar  properties \cite{mandelbrot}. 
Here, the power--law character of the distribution means that the largest part of the total wealth of the society is owned by a small fraction of the individuals, whereas most people only own a small fraction of it. For this reason, one also speaks of \emph{wealth condensation}  \cite{bouchaud} to indicate that a large amount of wealth is condensed in the hands of a small number of rich individuals.\\

Actually, in many cases Pareto's law only describes the \emph{tail} of a more complicated distribution. Indeed, as first noticed by Robert Gibrat  \cite{gibrat}, the small and middle ranges appear to be well described by a log--normal distribution. A random variable is said to be log--normally distributed if its logarithm is normally distributed. It follows that the log--normal distribution, also denoted \textbf{Gibrat's law}, has the form
\bea
\label{gibrat}
p(w)=\frac{\beta}{w\sqrt{\pi}}\exp\left[-\beta^2\log^2(w/w_0)\right]
\eea
where $\beta\equiv 1/\sqrt{2s^2}$ (\emph{Gibrat index}), $s^2$ is the variance and $w_0$ the mean of $\log(w)$. Differently from the power--law case, the log--normal distribution has a characteristic scale (even if, due to the logarithmic transformation, the distribution is much broader than a Gaussian with same mean and variance). On double--logarithmic axes, while Pareto's law looks like a straight line, Gibrat's law has a parabolic shape. 
Empirical studies show that $\beta$ varies in time and that $2\le\beta(t)\le 3$  \cite{souma1}. Therefore, in the most general case the value $w_*$ in (\ref{pareto}) has to be regarded as a crossover value marking the transition from the log--normal ($w<w_*$) to the power--law ($w>w_*$) regime. Importantly, this transition is non--smooth at the crossover value: the cumulative distributions have different slopes for $w<w_*$ and $w>w_*$  \cite{souma1}. As we shall discuss, this makes more difficult the possibility to reproduce empirical data by means of a simple model.\\

The above stylized facts of empirical distributions are well documented at the level of individuals  \cite{econowealth,yako,souma1}.
Similar results hold also for the size distribution of firms  \cite{sizes,simon}. 
It is also possible to analyse the economic system at the largest scale, with world countries as fundamental units and their gross domestic product (GDP) as their natural income  \cite{gdpgallegati,myalessandria,mytorino}. Again, one observes that the GDP distribution is characterized by a power--law--like right tail and by a different, log--normal--like left part. 
In summary, the above results suggest that many economic systems at different levels of aggregation are characterized by similar distributions of their associated wealth. The form of these distributions often ranges from a log--normal to a power--law behaviour, the most general case appearing a combination of both. 
Whether this actually allows one to speak of \emph{universal} distributions is of course an important but controversial point. Other fitting functions other than the purely log--normal and power--law forms have been proposed \cite{sizes,econowealth,yako}, and all of them offer advantages and drawbacks.
The survey of all possible statistical distributions is however beyond the scope of this work. In what follows, we shall mainly be interested in the microscopic origin of the observed patterns. This means that we shall review, and further explore, existing models focusing on the mechanisms of wealth dynamics that may be responsible for the heterogeneous shape of wealth distributions. To this end, the power--law and log--normal densities remain useful benchmarks capturing the essential difference between scale--free and single--scale wealth distributions.

\section{Theoretical approaches: independent versus interactive models}\label{sec}
Several microscopic models of wealth dynamics have been proposed. The simplest one, first proposed by Gibrat  \cite{gibrat} under the name of \emph{rule of proportionate growth}, is based on the assumption that the wealth $w_i$ of each individual $i$ evolves in time following a purely multiplicative stochastic process, that is
\be
\label{multiplicative}
\dot{w}_i(t)=\eta_i(t)w_i(t)
\ee
where the $\eta_i(t)$'s are independent random variables drawn from the same distribution, with assumed finite mean and variance. To be meaningful, the above stochastic differential equation has to be interpreted in a specified sense, following either Ito's or Stratonovich's convention  \cite{vankampen} (in the following sections, we shall always interpret equations of this type in the Stratonovich sense). Whatever the choice, it is clear that in the discretized form of (\ref{multiplicative}) $\log{w_i(t+1)}$ can be expressed as the sum $\log\eta_i(t)+\log\eta_i(t-1)+\dots$ of logarithms of the stochastic variable $\eta$. Then, as follows from the Central Limit Theorem, the logarithm of the wealth
approaches a normal distribution. This directly implies that the wealth $w_i$ is log--normally distributed according to (\ref{gibrat}).
Therefore the purely multiplicative stochastic model explains the appearance of Gibrat's law, but cannot reproduce the power--law tails of empirical wealth distributions. \\

A possible approach is to modify the stochastic equation (\ref{multiplicative}) in order to obtain different long--term forms of the wealth distribution. For example, the multiplicative process plus an additive noise term
\be
\label{additive}
\dot{w}_i(t)=\eta_i(t)w_i(t)+\xi_i(t)
\ee
can be shown to display (as long as $\langle\log\eta_i(t)\rangle<0$) an equilibrium solution for $p(w)$ having a power--law tail  \cite{sornette}. The Pareto index $\alpha$ is given by the condition $\langle\eta_i(t)^\alpha\rangle=1$, independently of the distribution of the additive term $\xi_i(t)$. With the same condition on $\langle\eta_i(t)\rangle$, a power--law distribution is also observed in the purely multiplicative model (\ref{multiplicative}) by imposing that $w$ cannot go below a cut--off value $w_{min}>0$  \cite{sornette}. While these models can reproduce the Pareto region (\ref{pareto}) of the empirical distribution, they do not reproduce Gibrat's law (\ref{gibrat}) for the lower wealth range.\\

One could then proceed in choosing other forms of the stochastic equation governing the individual wealth evolution, however there is a different approach focusing on the interactions taking place in the economy. Processes like (\ref{multiplicative}) and (\ref{additive}) rely on the assumption that the evolution of the wealth $w_i$ of an individual $i$ is uncoupled to the wealth $w_j$ of any other individual $j$, which is clearly an unrealistic hypothesis. By contrast, allowing the wealth to flow from an individual to another through transactions yields interactive, agent--based models.
Several such models have been proposed, most of them invoking either energy--like exchange mechanisms  \cite{econowealth,yako} or multi--agent stochastic differential equations  \cite{econowealth,bouchaud,tomaso}. The latter are our main interest here, since as we shall see they allow the introduction of an explicit interaction network defined among the agents.
In particular, we focus on the model proposed by Bouchaud and M\'ezard  \cite{bouchaud}, where the wealth of $N$ agents is governed by the following equation:
\be
\label{BM}
\dot{w}_i(t)=\eta_i(t)w_i(t)+\sum_{j\ne i}J_{ij}w_j(t)-\sum_{j\ne i}J_{ji}w_i(t)
\ee
where $w_i(t)$ is the wealth of agent $i$ at time $t$, the $\eta_i(t)$'s are independent Gaussian variables of mean $m$ and variance $2\sigma^2$ (accounting for random speculative trading such as market investments) and $J_{ij}$ is the element of an interaction matrix describing the fraction of agent $j$'s wealth flowing into agent $i$'s wealth (due to transactions between $i$ and $j$). Note the important property of invariance under wealth rescaling $w\to\lambda w$, which is required since money units are arbitrary  \cite{bouchaud}. If each agent exchanges wealth with every other, then $J_{ij}>0$ for each $i\ne j$. By further assuming that the exchanged fraction of wealth is the same for each pair of agents, we can set $J_{ij}=J/N$ where $J$ is a constant determining the strenght of the interaction. With this choice, eq. (\ref{BM}) reads 
\be
\label{full}
\dot{w}_i(t)=[\eta_i(t)-J]w_i(t)+J\langle w(t)\rangle
\ee
where $\langle w\rangle=\sum_i w_i/N$. It is possible to obtain analytically  \cite{bouchaud} the form of the wealth distribution $p(\tilde{w})$ (expressed in terms of the normalized wealth $\tilde{w}_i\equiv w_i/\langle w\rangle$):
\be
\label{eq}
p(\tilde{w})=\frac{(\alpha-1)^\alpha}{\Gamma[\alpha]}\frac{\exp[(1-\alpha)/\tilde{w}]}{\tilde{w}^{1+\alpha}}
\ee
with $\alpha\equiv 1+J/\sigma^2$. 
The exponential cut--off for low values of $w$ is essentially a smoothed form for the threshold $w^*$ appearing in eq.(\ref{pareto}).
Therefore, this model succeeds in reproducing Pareto's law, but the small and middle wealth ranges governed by the log--normal distribution (\ref{gibrat}) are again not reproduced, at least with the above choice of the matrix elements $J_{ij}$. However, as we clarify below, different forms of the interaction matrix can result in very different behaviours of the wealth distribution.

\section{Transaction networks}
In what follows, we explore the dependence of $p(w)$ on the interactions among agents in the Bouchaud--M\'ezard (hereafter BM) model. In particular, we consider a network where agents are represented by vertices and links correspond to transactions. The topology is fully specified by the \emph{adjacency matrix} $a_{ij}$, whose elements are $a_{ij}=1$ if there is a link from $i$ to $j$ and $a_{ij}=0$ otherwise. We shall only consider undirected networks for which $a_{ij}=a_{ji}$. Hence, the basic topological property characterizing each vertex is its number of connections or \emph{degree}, expressed as $k_i=\sum_j a_{ij}$. The statistical distribution of the degrees of all vertices is denoted by $P(k)$. If the interaction takes place only between directly connected agents, in the general case one can set $J_{ij}=(J/N)a_{ji}$.\\

In the purely multiplicative model (\ref{multiplicative}) agents are treated as independent, and this can be viewed as the extreme case of the BM model (\ref{BM}) when $J_{ij}=0$ for each pair of agents $i,j$. This corresponds to a trivial network with $N$ vertices and no edge, and yields the log--normal distribution (\ref{gibrat}). At the opposite extreme, the BM model with $J_{ij}=J/N$ for each $i,j$ corresponds to a fully connected network and yields the distribution (\ref{eq}) displaying power--law tails. This is the \emph{mean--field} case where each agent is subject to the same average influence of all other agents. It is then clear that different topologies result in different distributions. While the two extreme cases discussed so far are clearly unrealistic, it is interesting to explore intermediate, non--trivial topologies that can account for the observed mixed form of $p(w)$. This possibility has been explored in refs. 
 \cite{bouchaud,souma2,souma3,mypavia}. Indeed, it has been shown that a rich variety of outcomes can derive from the simple model (\ref{BM}). We briefly review these results below.

\section{Random, small--world and scale--free networks}
The simplest network model is the \textbf{random graph} proposed by Erd\"os and R\'enyi  \cite{erdos}, where each pair of vertices is connected by a link with probability $p$. The resulting degree distribution $P(k)$ has a Poissonian form  \cite{caldarelli,erdos}. 
Despite its simplicity, the random graph is particularly instructive when the BM model is defined on it. Since the extreme cases of independent and fully connected agents are recovered when $p=0$ and $p=1$ respectively, one could suspect that for intermediate values of $p$ there is a region displaying a combination of the log--normal and power--law form of the wealth distribution, similar to the empirically observed one. However, one can show  \cite{bouchaud} that, for large values of $p$, the behaviour is similar to that of fully connected networks, with a smaller value of the Pareto index $\alpha$. For small values of $p$, Gibrat's law still holds as in the case $p=0$.
In any case, no mixed form of $p(w)$ is observed.\\

Another simple network model is the \textbf{small--world} model proposed by Watts and Strogatz  \cite{watts}. One starts with a regular $d$--dimensional lattice (for example a ring) where each vertex is connected to its first $q$ neighbours and, with probability $p$, \emph{rewires} one end of each (undirected) link to a new randomly chosen vertex. The total number of links (and thus the mean degree $\langle k\rangle$) is then unchanged. Clearly, for $p=0$ one has a regular structure (with all vertices having the same degree), while for $p=1$ each link is randomly rewired, and the resulting network is not different from the random case (the degree distribution becomes very similar to a Poissonian one  \cite{caldarelli}). 
The BM model on regular rings and small--world networks with $d=1$ has been studied in refs.  \cite{souma2,mypavia} (however sometimes the different definition $J_{ij}=Ja_{ji}/k_i$ is used). It is found that when $p=0$ (regular ring), the form of $p(w)$ depends on the vertex degree $k=2q$: for small $k$ log--normal distributions are observed, while for large $k$ Pareto tails appear. This again suggests that, as in random networks (which here correspond to $p=1$), increasing the number of links promotes the transition from the log--normal to the power--law distribution in regular networks too ($p=0$). The most interesting result is that, in the non--trivial case of intermediate values of $p$ and $\langle k\rangle$, there exists a region where the wealth distribution can be fitted by a combination of the log--normal (for small $w$) and power--law (large $w$) forms  \cite{souma2}. This is encouraging, since these results get closer to the behaviour of real data. However the simulated distribution changes its shape smoothly, and this is not the most general pattern observed, as discussed before.\\

A model which is specifically intended to generate networks with power--law degree distributions is the \textbf{scale--free} model proposed by Barab\'asi and Albert  \cite{BA}. This is an evolving model starting with $m_0$ vertices and no link. At each time step, a new vertex is introduced and linked to $m$ pre--existing vertices chosen with probability proportional to their degree. 
High--degree vertices attract more and more links, and after enough time the degree distribution approaches a power law with exponent $\gamma=-3$. The BM model built on scale--free networks generated by this \emph{preferential attachment} mechanism has been studied in refs.  \cite{souma3,mypavia}. Once again, it is found that the wealth distribution is either log--normal or power--law, for small and large values of the mean degree $\langle k\rangle=2m$ respectively, with no mixed form occurring in the intermediate region.

\section{The role of heterogeneity}
In all the above models a transition from the log--normal to the power--law form of $p(w)$ occurs as the number of links increases. Only in non--trivial small--world networks there is an intermediate region displaying a mixed (however smooth) behaviour where both distributions are observed simultaneously. In the other cases the form of $p(w)$ changes quite abruptly from one form to the other as the value of the control parameter is changed.
A correct interpretation of these results is important.
Since in the above models the power--law character appears to be related to a larger average degree, and the log--normal behaviour to a smaller average degree, one could suspect that the simultaneous presence of low-- and high--degree vertices would determine the desired mixed form of the wealth distribution.
However, this hypothesis is contradicted by the observation that in the scale--free model vertices with large and small degree always coexist, but this never results in a mixed distribution. This means that the first--order properties alone, such as the degree distribution $P(k)$, are not enough to trigger the onset of the interesting pattern. As we show below, higher--order properties are required to enrich the structure of the wealth distribution. Interestingly, this is not a general feature of models based on stochastic differential equation: in the presence of additive rather than multiplicative noise, the degree distribution has a major effect on the form of the resulting wealth distribution  \cite{tomaso}.\\
\begin{figure}
\begin{center}
\includegraphics[width=.6\textwidth]{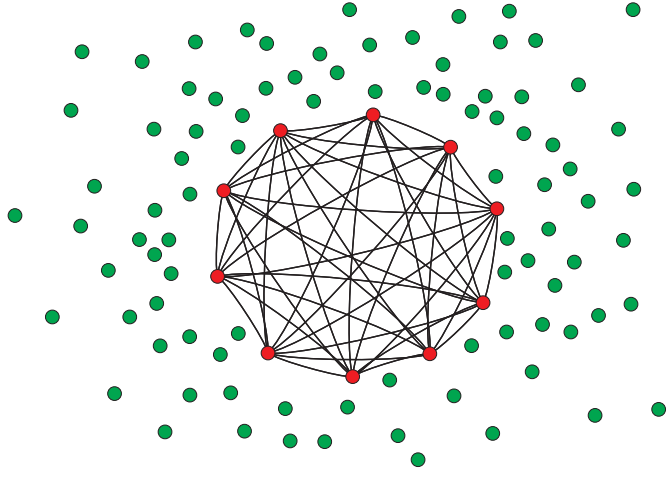}
\end{center}
\caption[]{Example of a mixed network with a fully connected core of $M=10$ vertices and $N-M$ completely isolated ones (figure produced using the Pajek software).}
\label{fig_mixednet}
\end{figure}

\begin{figure}
\begin{center}
\includegraphics[width=.6\textwidth]{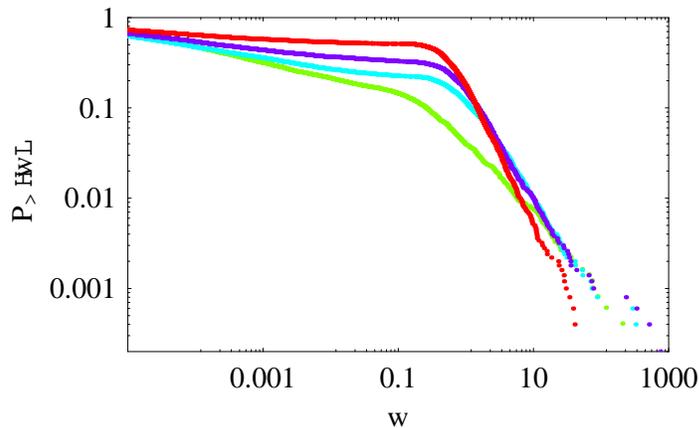}
\end{center}
\caption[]{Cumulative wealth distribution $P_>(w)$ for the BM model on a mixed network for different values of $M/N$: from top to bottom, $M/N=1/2,1/4,1/8,1/16$. In all cases $N=5000$, $J=\sigma^2=0.05$ and $m=1$. The wealth is rescaled to its average.}
\label{fig_mixed}
\end{figure}

We now present an example showing how the mixed form of $p(w)$ can be obtained in a trivial but instructive way  \cite{mypavia}. Consider an undirected network with $N$ vertices, $M$ of which are arranged in a fully connected graph and the remaining $N-M$ are completely isolated (clearly, $k_i=M-1$ for $i=1,\dots M$ and $k_i=0$ for $i=M+1,\dots N$). 
The evolution equation (\ref{BM}) then reduces to the mean--field case (\ref{full}) for the $M$ connected vertices, yielding the power--law disribution $p_1(w)$, and to the independent agent case (\ref{full}) for the $N-M$ isolated vertices, yielding the log--normal distribution $p_2(w)$. As a consequence, the global distribution $p(w)$ is such that the total number $Np(w)$ of agents with wealth $w$ equals the number $Mp_1(w)$ of the connected ones with wealth $w$ plus the number $(N-M)p_2(w)$ of the isolated ones with wealth $w$. In other words
\be
\label{mixing}
p(w)=\frac{M}{N}p_1(w)+\left(1-\frac{M}{N}\right)p_2(w)
\ee
and the only control parameter is $M/N$  \cite{mypavia}. In fig. \ref{fig_mixed} we show $p(w)$ obtained by means of numerical simulations for various choices of $M/N$. The observed form is clearly the sum of the contributions coming from the two sets of vertices.\\

\begin{figure}
\begin{center}
\includegraphics[width=.6\textwidth]{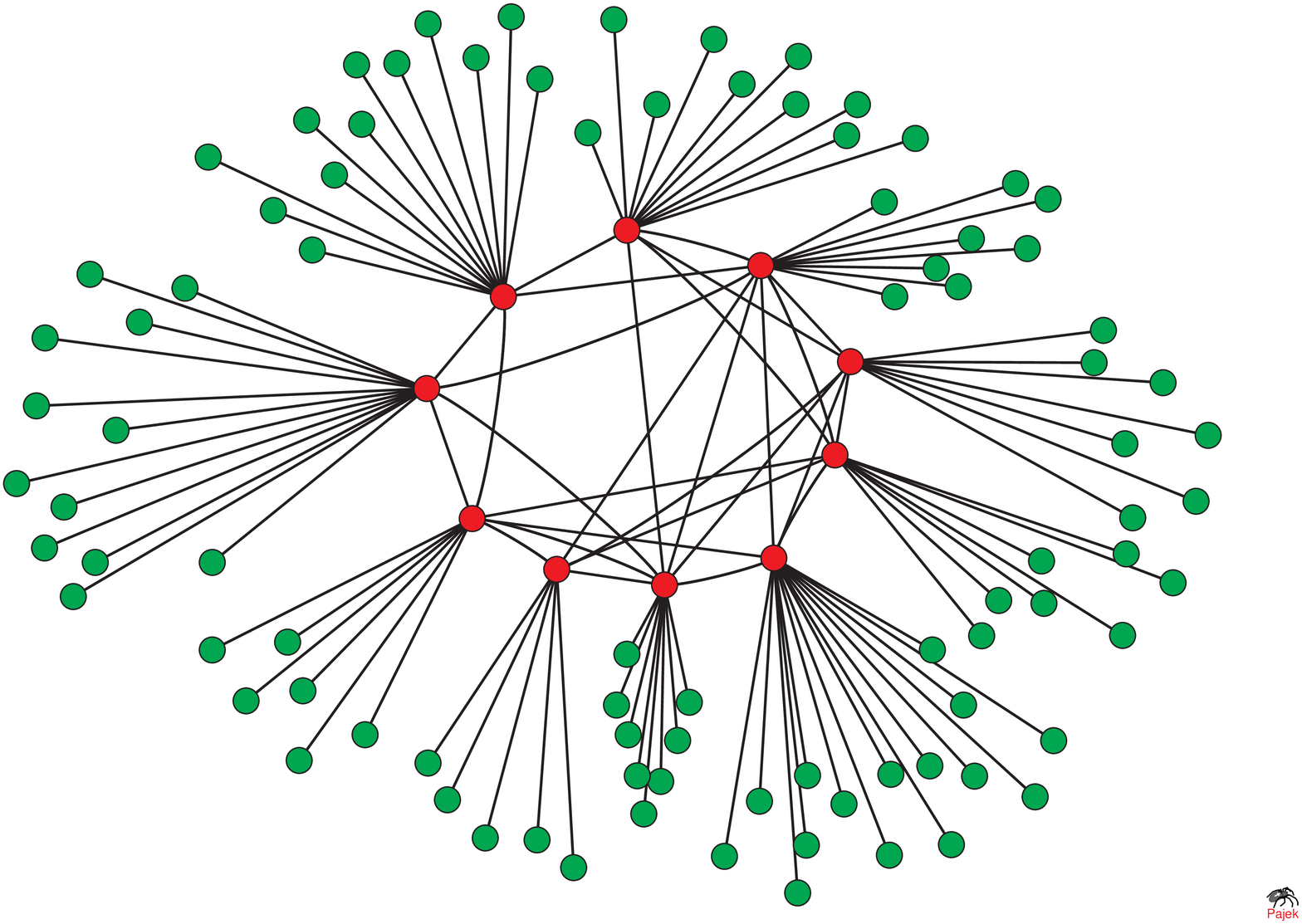}
\end{center}
\caption[]{Example of an octopus network with a core of $M=10$ randomly connected vertices and $N-M=90$ tentacles with single connections to the core (figure produced using the Pajek software).}
\label{fig_octopusnet}
\end{figure}

\begin{figure}
\begin{center}
\includegraphics[width=.6\textwidth]{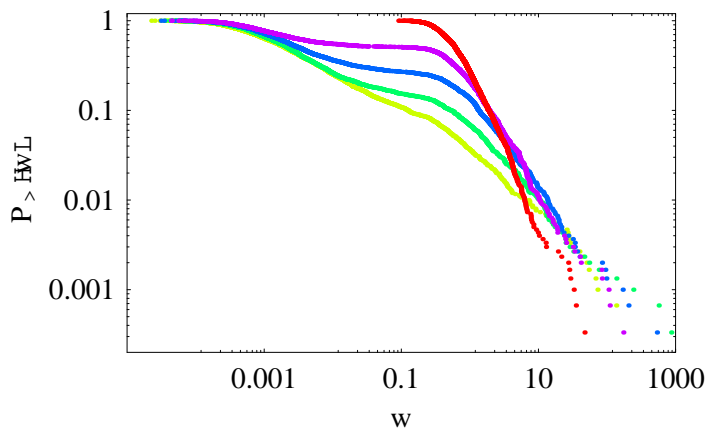}
\end{center}
\caption[]{Cumulative wealth distribution $P_>(w)$ for the BM model on an octopus network for different values of $M/N$: from top to bottom, $M/N=1,1/2,1/4,1/8,1/16$. In all cases $N=3000$, $J=\sigma^2=0.05$ and $m=1$. The wealth is rescaled to its average.}
\label{fig_octopus}
\end{figure}

The above extreme example suggests that the mixed character of the empirically observed wealth distribution might be the effect of the simultaneous presence in the network of regions with different link density, either well or poorly connected. In order to further explore this possibility, one needs to check different topologies having the same key ingredient.
Indeed, the mixed behaviour captured by eq.(\ref{mixing}) is still found if the $N-M$ vertices are arranged in a periodic chain, as follows directly from the results we reported on the log--normality of $p(w)$ for the regular ring.
As an additional testbed, endowed with a richer structure that allows further analyses, we consider an \emph{octopus} network (see fig. \ref{fig_octopusnet}) where $M$ vertices are connected in a random network forming a denser core and each of the remaining $N-M$ vertices has only one connection (looking like a \emph{tentacle}) to a randomly chosen vertex in the core. 
In this case too, the mixed behaviour is observed  \cite{mypavia}, as shown in fig. \ref{fig_octopus}. Indeed, the simulated distributions look much like the real distributions  \cite{econowealth}.
The results for the octopus topology also inform us that for the mixed behaviour to appear we do not need the network to display disconnected regions as in the preceding example of fig. \ref{fig_mixednet}. The fundamental ingredient appears to be the coexistence of regions with high (the core) and low (the periphery) link density. We then interpret that in one--dimensional small--world networks the mixed shape is obtained because, for not too large $\langle k\rangle$ and for suitable values of the rewiring probability $p$, in the network a set of randomly connected vertices and a part of the original ring coexist. We then expect the latter to give a log--normal contribution to $p(w)$, and the former to introduce the power--law tail. 

\section{Assortativity and correlations}
\begin{figure}
\begin{center}
\includegraphics[width=.3\textwidth]{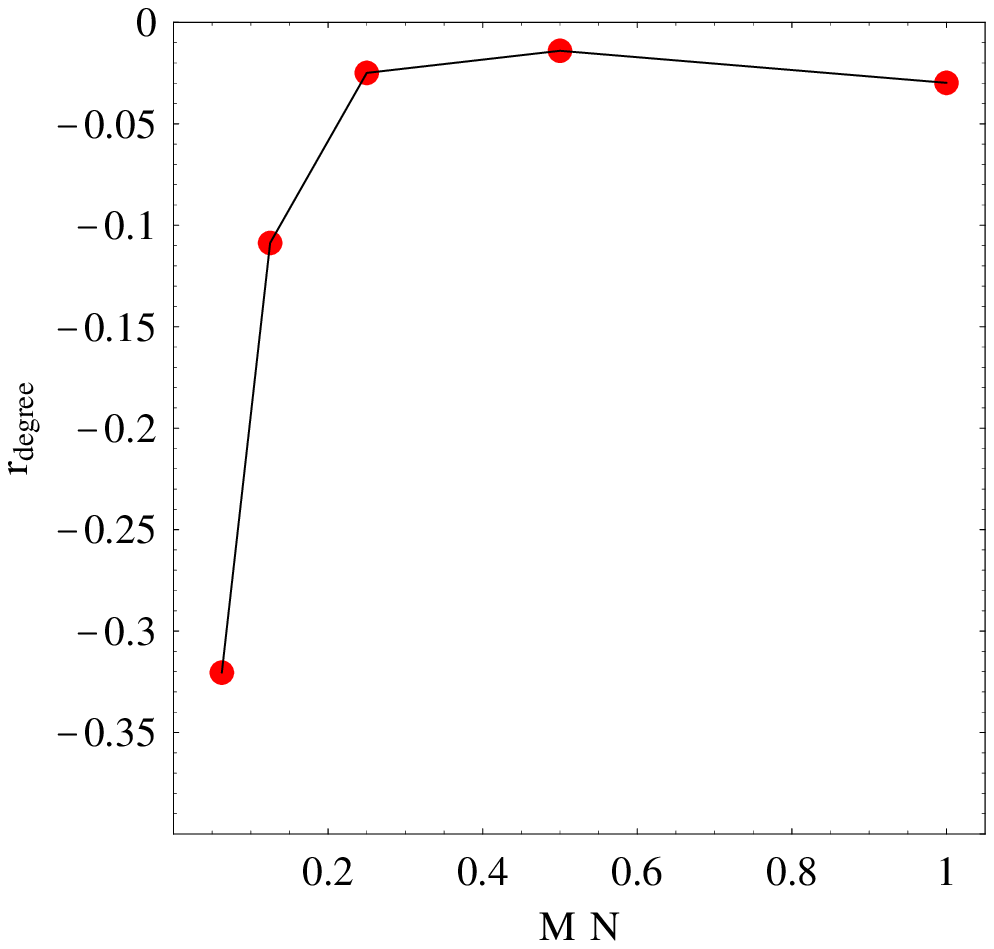}
\includegraphics[width=.3\textwidth]{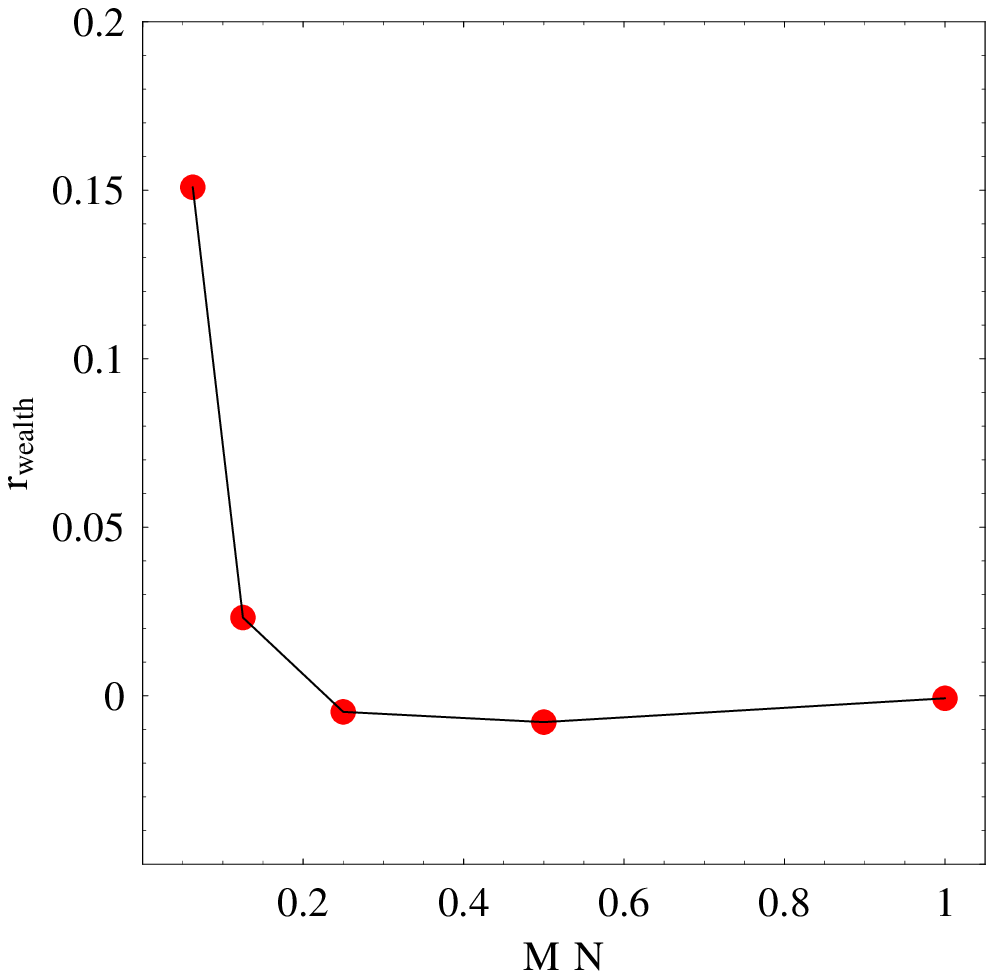}
\includegraphics[width=.3\textwidth]{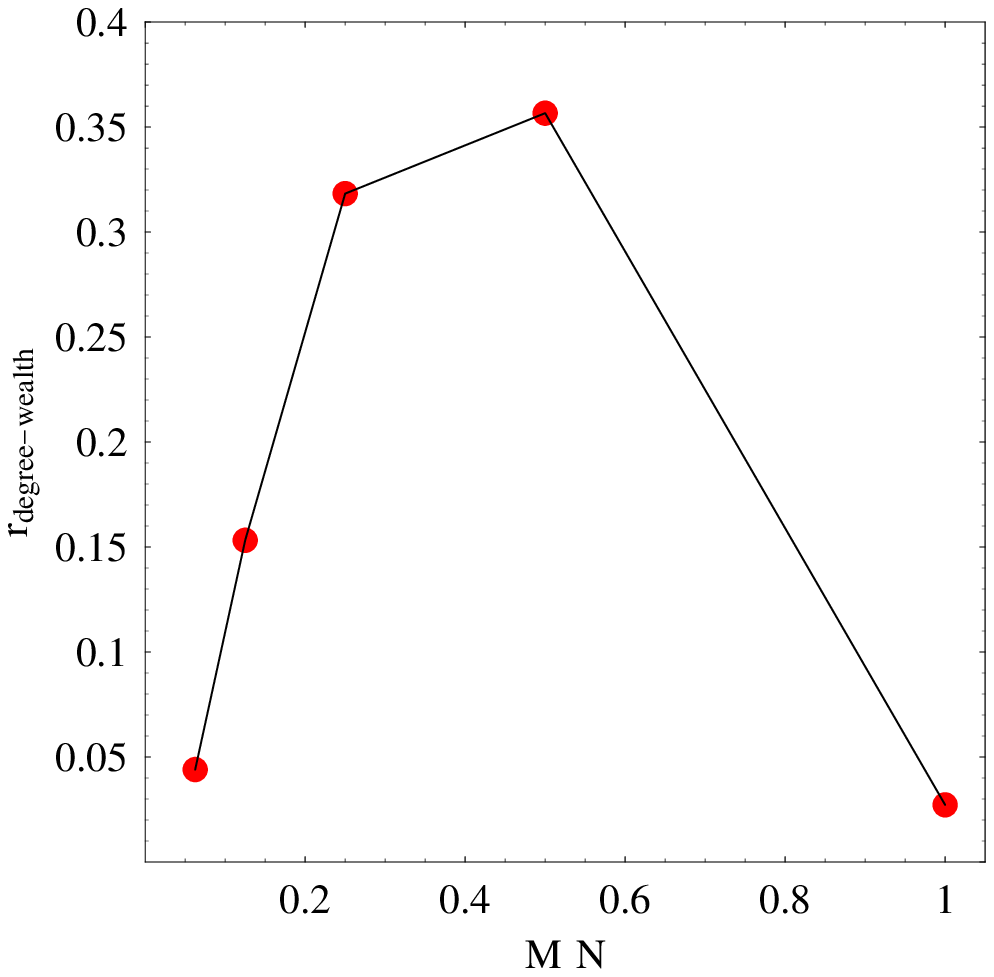}
\end{center}
\caption[]{Correlations coefficients for the BM model on an octopus network for different values of $M/N$. In all cases $N=3000$, $J=\sigma^2=0.05$ and $m=1$.}
\label{fig_r}
\end{figure}

Since the degree distribution alone does not allow to fully characterize the stationary state that the BM model gives rise to, here we inspect higher--order topological properties. In particular, we study the assortativity  \cite{newman1,newman2} of the network, which is a measure of the pairwise correlations between the degrees of neighbouring vertices. 
A means to quantify this property is given by the coefficient of \emph{assortativity by degree} \cite{newman1,newman2} defined as
\begin{equation}
r_{degree}\equiv\frac{1}{\sigma_q^2}\sum_{jk}jk(e_{jk}-q_j q_k)
\label{eq_rdegree}
\end{equation}
where $e_{jk}$ is the fraction of links between vertices with degrees $j$ and $k$, $q_k=\sum_j e_{jk}$ is the probability that a randomly chosen link leads to a vertex with degree $k$, and $\sigma_q^2$ is the variance of this distribution. Being a Pearson correlation coefficient, $r_{degree}$ ranges between $-1$ and $1$. Positive values indicate a tendency for vertices with large degree to connect more frequently with each other (assortativity), while negative values indicate a tendency for large--degree vertices to connect with low--degree ones (disassortativity). The uncorrelated case corresponds to $r_{degree}=0$. The Erd\"os--R\'enyi random graph and the Barab\'asi--Albert model defined above are two examples of uncorrelated networks  \cite{newman1}.
By contrast, the octopus network is clearly disassortative, since the peripheral unit--degree vertices are connected with the large--degree vertices in the core. The values of $r_{degree}$ for the octopus network are displayed in fig.\ref{fig_r} for various choices of the parameter M/N. The disassortative character increases as the fraction of peripheral vertices increases, since the difference between the peripheral degrees and the core degrees increases.
This result identifies the pairwise topological correlations as a first sufficient ingredient for the non--smooth behaviour of $p(w)$ to appear.\\

It is also important to investigate how the topology relates to higher--order properties of the resulting wealth distribution. That is, one could add to the overall knowledge given by $p(w)$ the information regarding the correlation between the wealth of neighbouring vertices. As a straightforward measure of this property we define the \emph{assortativity by wealth}
\begin{equation}
r_{wealth}\equiv\frac{1}{\sigma_p^2}\int wv(e_{wv}-p_w p_v)dwd v
\end{equation}
where, in analogy with eq.(\ref{eq_rdegree}), $e_{wv}$ is the fraction of links between vertices with (rescaled) wealth in the range $[w,w+dw]$ and vertices with wealth in the range $[v,v+dv]$. Here $p_w=\int e_{wv}dv$ and $\sigma_p^2$ is the variance of the distribution $p_w$. 
The value of $r_{wealth}$ computed on the long--term state of the BM model on the octopus network are shown in fig.\ref{fig_r}. We find that $r_{wealth}$ follows a trend opposite to that of $r_{degree}$: as $M/N$ decreases, the wealth of neighbouring vertices is more and more positively correlated.\\

Finally, we also investigate the degree--wealth correlation $r_{degree-wealth}$ that we simply define as the Pearson correlation coefficient between the degree and wealth of each vertex. As shown in fig.\ref{fig_r}, we find that when $r_{degree}$ and $r_{wealth}$ are significantly different from zero (small $M/N$), $r_{degree-wealth}$ is very low. This means that the degree of a vertex does not correlate with its wealth. This is in accordance with the fact that in this region one has $r_{degree}<0$ and $r_{wealth}>0$ simultaneously, which is not possible if degree and wealth are positively correlated. 
By contrast, for large values of $M/N$ $r_{degree-wealth}$ becomes large, signalling a strong correlation between wealth and degree (except for $M/N=1$, where all vertices have nearly the same degree while their wealth is broadly distributed, resulting in a vanishing $r_{degree-wealth}$). Interestingly, when this occurs $r_{degree}$ and $r_{wealth}$ are negligible.\\

Therefore we find that, as $M/N$ increases, the system changes from a state where degree and wealth do not correlate,  but where there is strong disassortativity by degree and assortativity by wealth, to a different one where degree and wealth are positively correlated, but the assortativity (either by wealth or degree) is negligible.  These considerations allow a deeper understanding of the interplay between topology and dynamics, which is not captured by the wealth distribution alone.

\section{Conclusions}
We discussed how the empirically observed forms of wealth distributions can be reproduced by a single stochastic model of wealth dynamics. The long--term shape of the distribution strongly depends on the topology of the transaction networks among economic units. The purely log--normal and power--law forms arise quite naturally if the network diplays a homogeneous density of links. By contrast, the frequently observed mixed shape appears to be related to a heterogeneous link density, which we traced back to the presence of a core region in the network. We have therefore concluded that the first--order topological properties alone (such as the degree distribution) are not sufficient to specify the dynamical outcome of the process, since higher--order correlations play a major role. In order to characterize these effects, we have studied the impact of assortativity on the onset of nontrivial wealth distributions and wealth correlations.
Interestingly, the type of heterogeneity we have introduced appears to be widespread in real networks, as a result of strong community structure  \cite{caldarelli}, rich--club ordering  \cite{richclub} or assortative mixing  \cite{newman1,newman2}. Therefore the effects considered here represent a prototype for the dynamical processes that are expected to take place on real economic networks.

\end{document}